# Principle components analysis for seizures prediction using wavelet transform

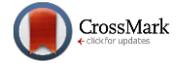

Syed Muhammad Usman [1, *], Shahzad Latif [1], Arshad Beg [2]

[1]Shaheed Zulfikar Ali Bhutto Institute of Science and Technology, Islamabad, Pakistan
[2]Department of Electrical Engineering, FAST National University of Computer and Emerging Sciences, Faisalabad Campus, Pakistan

A B S T R A C T

Epilepsy is a disease in which frequent seizures occur due to abnormal activity of neurons. Patients affected by this disease can be treated with the help of medicines or surgical procedures. However, both of these methods are not quite useful. The only method to treat epilepsy patients effectively is to predict the seizure before its onset. It has been observed that abnormal activity in the brain signals starts before the occurrence of seizure known as the preictal state. Many researchers have proposed machine learning models for prediction of epileptic seizures by detecting the start of preictal state. However, pre-processing, feature extraction and classification remains a great challenge in the prediction of preictal state. Therefore, we propose a model that uses common spatial pattern filtering and wavelet transform for preprocessing, principal component analysis for feature extraction and support vector machines for detecting preictal state. We have applied our model on 23 subjects and an average sensitivity of 93.1% has been observed for 84 seizures.

## 1. Introduction

Epilepsy is a common disease that is caused due to abnormal activity of neurons in brain (Thurman et al., 2016). Epilepsy patients undergo frequent unexpected seizures. More than 1 % of the world's population is suffering from this disease. Brain activity of such patients can be divided into multiple states. The state during which seizure occurs is known as ictal state (Feldwisch-Drentrup et al., 2011). An abrupt change occurs in brain activity during ictal state that leads towards loss of consciousness of the patient. Epilepsy can be treated by medicines (Schmidt and Schachter, 2014) and sometimes surgical treatments (Van Buren, 1987) are required to remove the tissues that cause epilepsy. However, both these methods are not very efficient as medicines do not prevent the seizure and surgical procedure is invasive method and is only effective when tissues of certain portion of brain are causing seizures. If seizures are not controlled, then the patient's life is badly affected in terms of social, jobs and family. In fact epilepsy patient cannot live a good life.

Therefore, a better option to treat epilepsy is to predict the seizures and prevent it with the help of medicines. Epileptic seizures can be divided into four states i.e., interictal state, preictal state (Le Van Quyen et al., 2005), ictal and post-ictal stats. Preictal state starts few minutes before the start of seizure. Predicting epileptic seizure before it actually occurs is itself not an easy task. Various researchers have applied different algorithms for prediction of epileptic seizures in past. Quantitative studies have shown that preictal state is quite useful for prediction of epileptic seizure. This state may start 10 to 30 minutes or in some case more than 30 minutes. Lyapunov exponents (Blanco et al., 1995) are useful in differentiating between these states. It is quite evident by analyzing multiple EEG recordings that there is a significant change that occurs in all the state of epileptic seizure. It has been observed that noise added during recording of EEG signals by placing electrodes on patient's scalp. Therefore, an extensive preprocessing is required to remove the noise from EEG recordings. Another problem in prediction is feature selection for classification of multiple states of seizure. Hybrid features selection methods has been used in for classification. Principal component analysis (PCA) (Ghosh-Dastidar et al., 2008) is applied in for preprocessing of EEG signals. A significant variability in hear rate of patients occurs during seizure. Advanced empirical mode decomposition (Flandrin et al., 2004) is also used with Morlet wavelet transform to predict epileptic seizures.

* Corresponding Author.
Email Address: Muhammad.usman@szabist-isb.edu.pk (S. M. Usman)
https://doi.org/10.21833/ijaas.2019.03.008
Corresponding author's ORCID profile:
https://orcid.org/0000-0002-0504-3558

Heart rate variability features and with multivariate statistical process control gives better accuracy for prediction of epileptic seizures. Many researchers have predicted epileptic seizures on the basis of multiple linear features and non-linear features. Fig. 1 shows multiple channels of EEG signals.

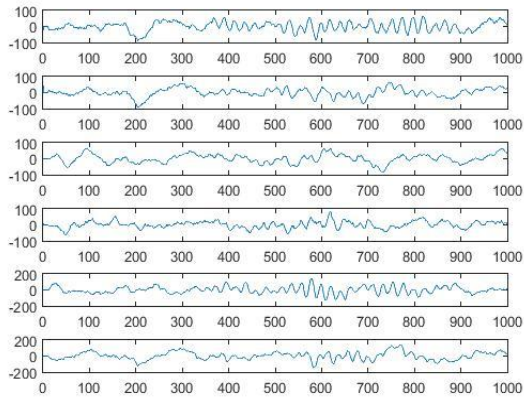

**Fig. 1:** Multiple channels of EEG signals

These features include relative power, approximate entropy (Ocak, 2009), Lyapunov exponents, complexity, Hjorth parameters (Obermaier et al., 2001) and multiple statistical features in both time and frequency domain. Wavelet transform (Hazarika et al., 1997) is considered a powerful tool for preprocessing of EEG signals. Low frequency components can be extracted from EEG signals. If we are able to do effective pre-processing and select a suitable model after feature extraction for classification of preictal and interictal state then prediction of epileptic seizure will be very useful in health sector.

## 2. Related works

Seizure prediction is quite useful as it is the only method to prevent the seizure before it actually happens. Therefore various researchers have proposed different algorithms and methods for prediction of seizures with high accuracy and sensitivity. Senger and Tetzlaff (2016) have applied principle component analysis (PCA) for preprocessing of EEG signals and then zero crossing levels have been observed and noted for prediction of epileptic seizure. Parvez and Paul (2016) has proposed that classification between preictal and interictal state for prediction of epileptic seizures need pre-processing as well as post processing of EEG signals. For pre-processing differential window has been applied on EEG signals to make them more distinct for classification. After applying differential window as pre-processing, multiple features have been extracted using phase correlation. Support vector machines (SVM) has been applied for classification between interictal and preictal states. For post-processing k out of n is applied for getting better prediction results. Approximately 91.95% accuracy has been achieved on dataset acquired by university of Freiburg.

Shiao et al. (2017) proposed a system to predict epileptic seizure using SVM. Test samples have been passed to classifier for 1 hour recordings and 20 seconds non-overlapping window. Pre-ictal state has been assumed to be few minutes before the onset of the seizure. Post-processing is also done due to unbalanced data of preictal and interictal states. Majority voting has been used as a post processing step in order to classify state as either pre-ictal or interictal state.

Hosseini et al. (2017) have proposed a cloud based solution for prediction of epileptic seizures. Butterworth filter and notch filter have been applied as preprocessing step to remove noise and artifacts from the EEG signals. Wavelet transform is also applied in order to extract spikes of preictal state. After applying pre-processing, multiple time and frequency domain features have been extracted from the EEG signals including entropy, correlation coefficients, zero crossing, average power and statistical moments. Multiple techniques for classification between preictal and interictal state has been applied including linear and non-linear SVM, neural networks and convolution neural networks (CNN). CNN provides maximum classification accuracy and true positive rate. Parvez and Paul (2017) used phase correlation for feature extraction and least square support vector machine has been used for classification. U of v classification method has been applied as post processing for removing artifacts and to avoid misclassification. Prediction accuracy of 95% has been observed by applying this method on publically available dataset of 21 patients of intracranial EEG signals.

Wang and Lyu (2015) proposed a method of feature extraction using frequency and amplitude on epoch basis. Feature selection has been done using the efficiency of the features. SVM is used as classifier and more than 90 % sensitivity has been observed on the publically available free database.

In Teixeira et al. (2014), authors have extracted univariate features from non-overlapping window and then classification is done using support vector machines. Butterworth filter is also applied as pre-processing for noise removal in EEG signals. Teixeira et al. (2014) have extracted spectral features for classification between pre-ictal and interictal states. SVM is used as classifier and sensitivity of 75.8% has been observed in 87 seizures. In Huang et al. (1998), wavelet transform has been used for pre-processing. Authors have extracted energy and entropy after applying wavelet transform to EEG signals. Only few channels have been selected for six patients and sensitivity of 88% has been observed after classification.

Table 1 shows the comparison of multiple methods used for prediction of epileptic seizures by different researchers. Pre-processing, feature extraction and choosing suitable model for prediction are the main challenges in prediction.

**Table 1:** Comparison of existing models for epilepsy prediction

| Method | Subjects | Seizures | EEG channels | Features | Sensitivity |
|---|---|---|---|---|---|
| Senger and Tetzlaff (2016) | 20 | 103 | 20 | 05 | 71.55 |
| Teixeira et al. (2014) | 224 | 87 | 06 | 22 | 73.08 |
| Shiao et al. (2017) | 06 | 38 | 16 | 03 | 87.83 |
| Zandi et al. (2013) | 20 | 86 | 18/23 | 23 | 88.34 |
| Teixeira et al. (2014) | 24 | 87 | 06 | 12 | 73.98 |
| Brinkmann et al. (2016) | 07 | 08 | 16-24 | 10 | 88 |
| Rasekhi et al. (2013) | 10 | 86 | 6 | 22 | 71.97 |
| Gadhoumi et al. (2012) | 06 | 86 | 2/3 | 03 | 88 |

Therefore, we propose an effective method of pre-processing with the help of common spatial pattern filtering, noise removal by using wavelet transform and feature extraction using principle component analysis. Following section provides a brief detail of our proposed methodology.

## 3. Proposed method

The pre-ictal state is very useful for seizure prediction, as it starts few minutes before the seizure. This made it possible for us to be able to predict epileptic seizure, if we successfully detect the start of pre-ictal state. The aim of this research is to predict epileptic seizure, by detecting the start of pre-ictal state sufficient time, before the ictal state or onset of seizure starts. Early prediction (Huang et al., 1998) helps patients, as medication can be done by the doctors to prevent the seizure. Due to this medication, the patient can now perform his or her routine activities without any interference from seizures. After critical considerations of these states, we have proposed a model to detect the start of the pre-ictal state. However, EEG data acquisition by placing electrodes on the scalp of the patient is out of the scope of our research. Fig. 2 shows the flow chart of our proposed method.

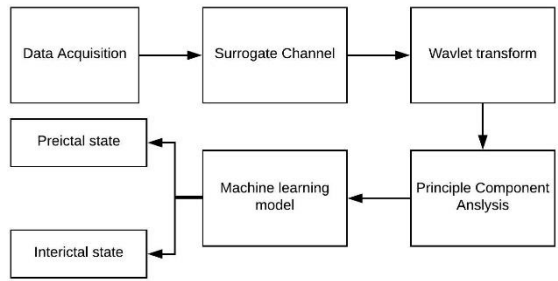

**Fig. 2:** Flow chart

In the first step, data acquisition is done. As data acquisition if not in the scope of this research, therefore we have used a publically free online available dataset of CHB-MIT (Moody et al., 2011). The dataset has been acquired by placing 23 electrodes on the scalp of 22 subjects. The data contains EEG recordings of 22 patients for many hours and for convenience each recording has been divided into 1 hour sessions. The data has been sampled at rate of 256 Hz. We have selected only those sessions where ictal state starts at least after 20 minutes of recordings. We have performed preprocessing of the data in two stages; in the first stage, 23 channels EEG signals are converted into a surrogate channel, which is a single signal to improve the SNR. After converting the multiple channels EEG signals into single channel surrogate signal, we have applied wavelet transform for de-noising of signal. Once the pre-processing is completed, we have applied principle component analysis to get 10 features from the dataset with a non-overlapping window having size of 5 seconds. Finally, classification is done using support vector machines.

### 3.1. Common spatial pattern filtering (CSP)

Common spatial pattern filter (CSP) (Ang et al., 2008) is used for converting multiple channels EEG signals into single surrogate channel EEG signal. For EEG signals, CSP performs much better as it increases variance between interictal and pre-ictal classes and also increase signal to noise ratio of EEG signals. Assume signals of two different states of preictal and interictal be X1 and X2. We can compute filter coefficients with the help of following equations.

$$R_1 = \frac{(X_1 X_1^t)}{trace(X_1 X_1^t)} \quad (1)$$

$$R_2 = \frac{(X_2 X_2^t)}{trace(X_2 X_2^t)} \quad (2)$$

$$R = R_1 + R_2 \quad (3)$$

$$[Evec, Eval] = eig(R) \quad (4)$$

R1 and R2 are the matrices obtained as a result of dividing squares of the signals of each state with the trace of square of signals. Both matrices are added to get a single resultant matrix R (Usman et al., 2017). Evec and Eval represent eigenvector and eigenvalues respectively. Now assume that a matrix D consists of diagnomal elements of eigenvector.

$$w = \sqrt{D^{-1}} Evec^t \quad (5)$$

$$S_1 = w R_1 w^t \quad (6)$$

$$S_2 = w R_2 w^t \quad (7)$$

$$[B, D] = eig(S_1, S_2) \quad (8)$$

$$Filter = B^t w \quad (9)$$

$S_1$ and $S_2$ are computed by multiplying with weight w to both signals $X_1$ and $X_2$. Eigenvector

decomposition is applied to $S_1$ and $S_2$. Filter coefficients are in the form of a 1x23 vector in our case. This filter is applied to 23 channels EEG signals to get a single surrogate channel EEG signal.

## 3.2. Principle component analysis (PCA)

Principle component analysis (PCA) (Ghosh-Dastidar et al., 2008) is a statistical method that transforms higher dimensional space into lower dimension features. Linear transformation is used to rotate coordinate system. The axes of new coordinate system known as components are formed after combining the original axes linearly. Principle component also known as primary axis is selected on the basis of maximum variance of data in particular direction. Secondary axis is orthogonal to principle component and it shows next highest variance of data. Similarly, other components are selected and in this way data is converted into lower dimensions. Upon applying PCA, most of variance of data is in first few components. As a result, only those components that have large variations are kept and rests are ignored. In the first step of PCA, n dimensional mean vector mu is computed then nxn covariance matrix R is computed and sorted in descending order or eigenvalues. Largest eigenvalues are chosen after sorting. Other dimensions are considered as noise. If we form a n x n matrix A having n eigenvectors in columns then the data after pre-processing can be obtained from the following equation.

$$X' = A^t(x - \mu) \quad (10)$$

## 3.3. Wavelet transform

Wavelet transform decomposes signals by using basis functions known as wavelets. These basis functions are obtained using mother wavelet or prototype wavelet by dilations, contractions and shifting the signals. Wavelet transform can be of two types, i.e.; continuous and discrete wavelets transform CWT and DWT respectively. In CWT signal is convolved with basis function of wavelet. However, in CWT data needs to be digitized. In discrete wavelet transform (Adeli et al., 2003), inner product of signal with the basis function. This basis function is known as wavelet function. Fig. 3 shows the original details coefficients obtained after applying Haar transform. Whereas, Fig. 4 shows the decomposition of signal after applying wavelet transform.

If $f(t)$ be function of time t and square integrals, then continuous wavelet transform is defined as:

$$W_{a,b} = \int_{-\infty}^{+\infty} f(t) \frac{1}{\sqrt{|a|}} \varphi * \left(\frac{t-b}{a}\right) dt \quad (11)$$

In above equation, a and b are real numbers, and * represents complex conjugation. Wavelet function or basis function can be defined as following:

$$\varphi_{a,b}(t) = \frac{1}{\sqrt{|a|}} \varphi\left(\frac{t-b}{a}\right) \quad (12)$$

Eq. 1 can be re-written as

$$W_{a,b} = \int_{-\infty}^{+\infty} f(t) \varphi_{a,b}(t) \quad (13)$$

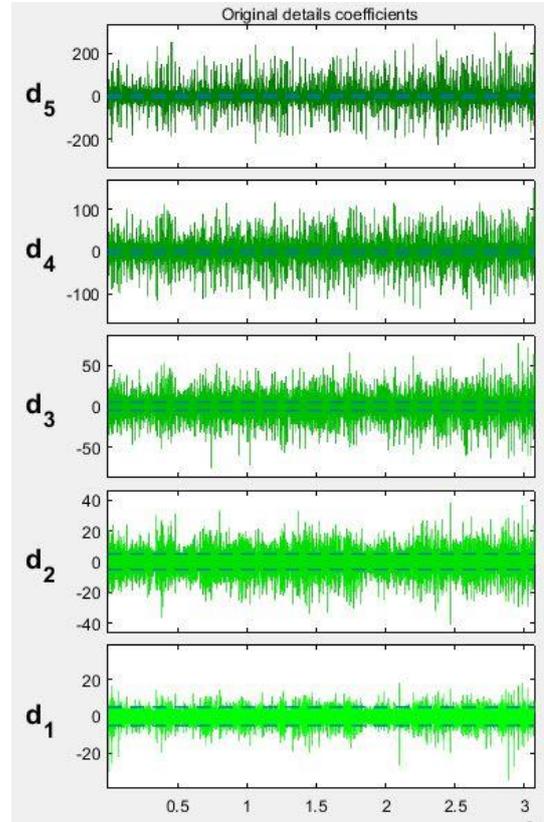

**Fig. 3:** Original details coefficients

We have applied Haar wavelet transform to our signal to de-noise the EEG signals.

We have compared multiple classifiers for classification including k-nearest neighbor classifier, Naïve Bayes and support vector machines. It has been concluded that only support vector machines performs better in terms of accuracy and true positive rate. Therefore, we have applied support vector machines for classification between interictal and pre-ictal states.

## 4. Results

We have applied our proposed model on publically available dataset of CHB-MIT having 22 subjects and multiple sessions for each patient. We have selected only those sessions that have at least 20 minutes time to start ictal state. Therefore, we have tested 84 sessions. An average sensitivity of 93.1 % has been observed by our proposed model. If our model is applied then we are able to predict epileptic seizure before it actually occurs and therefore, by doing medication it can be prevented that can avoid loss of patients in terms of health and social life. Fig. 5 shows comparison between sensitivity obtained with different methods and proposed method.

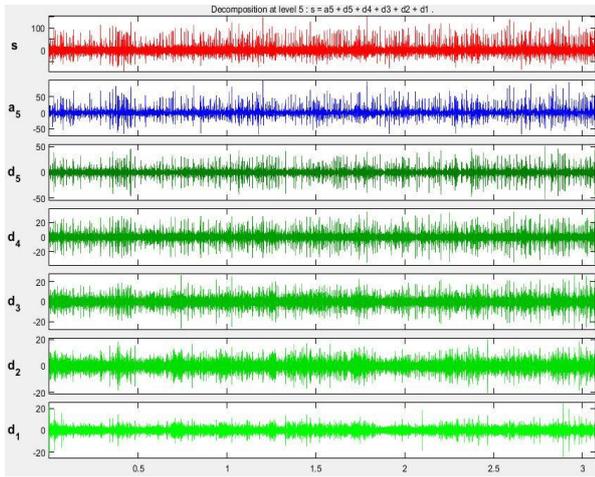

**Fig. 4:** Decomposition of signal using Haar wavelet

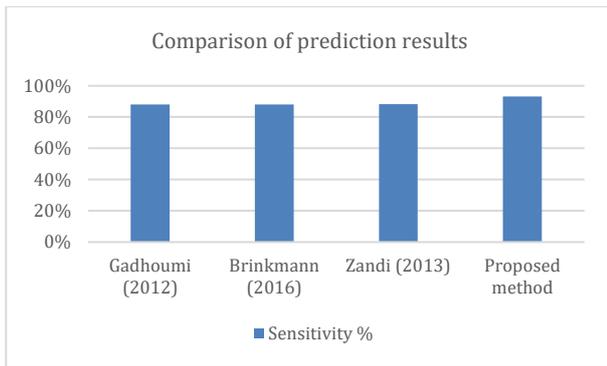

**Fig. 5:** Comparison of prediction results

## 5. Conclusion

It has been observed that effective way of dealing with epileptic seizure is to prevent the seizure by predicting it sufficient time before it actually occur. Our proposed model helps the epilepsy patients by prediction of seizures well before time with a greater true positive rate. However, in future we can increase the anticipation time and sensitivity by applying more techniques for preprocessing and feature extraction. As data acquisition is not in scope of our research, but in future we can also make an online system for prediction by doing data acquisition and prediction in real time online environment.

## Compliance with ethical standards

## Conflict of interest

The authors declare that they have no conflict of interest.